# Turbulence demonstrates height variations in closely spaced deep-sea mooring lines

by Hans van Haren

Royal Netherlands Institute for Sea Research (NIOZ), P.O. Box 59, 1790 AB Den Burg, the Netherlands.
e-mail: hans.van.haren@nioz.nl

**Abstract.** It may be important to precisely know heights of moored oceanographic instrumentation. For example, moorings can be closely spaced or accidentally be located on small rocks or in small gullies. Height variations O(1 m) will yield registration of different values when conditions such as small-scale density stratification vary strongly. Such little height variations may prove difficult to measure in the deep sea, requiring high-accuracy pressure sensors preferably on all instruments in a mooring-array. In this paper, an alternative method for relative height determination is presented using high-resolution temperature sensors moored on multiple densely-spaced lines in the deep Western Mediterranean. While it was anticipated that height variations between lines could be detected under near-homogeneous conditions via adiabatic lapse rate O(0.0001°C m$^{-1}$) by the 0.00003°C-noise-level sensors, such was prevented by the impossibility of properly correcting for short-term bias due to electronic drift. Instead, a satisfactory height determination was found during a period of relatively strong stratification and large turbulence activity. By band-pass filtering data of the highest-resolved turbulent motions across the strongest temperature gradient, significant height variations were detectable to within ±0.2 m.

# 1 Introduction

Height variations in moored oceanographic instrumentation can occur above unknown topographic features such as small rocks and gullies and, e.g., due to line stretching under several kN of buoyancy pull. If closely-spaced mooring lines and attached instrumentation are used, one may need to correct for the unknown height variations. Such a correction is possible using high-resolution temperature sensors.

In this technical-performance paper, deep-sea instrumental-height determinations will be demonstrated using 45 mooring lines of 125-m tall and 9.5-m apart horizontally each holding 65 self-contained high-resolution temperature 'T-'sensors in a 70-m diameter steel-pipe ring (van Haren et al., 2021). Each line was pulled up by a net buoyancy of 1.25 kN, imposed by a single buoy on top, and was attached to the anchoring ring via a suspended steel-cable grid. The ring was moored on a flat seafloor in nearly 2500-m deep weakly density-stratified waters of the Western Mediterranean Sea. The 'large-ring mooring' was constructed for three-dimensional studies of deep-sea internal waves, their breaking and turbulence generation, to learn more about their dynamical development via short movies (van Haren et al. 2026) and detailed statistics.

# 2 Materials and Methods

A nearly half-cubic-hectometer of seawater was sampled using 2925 self-contained updated high-resolution, stand-alone T-sensors, new version 'NIOZ4n'. The ensemble large-ring mooring (Fig. 1a) was deployed in drag-parachute controlled free-fall at the <1° flat and 2458-m deep seafloor of 42° 49.50′N, 006° 11.78′E of the Northwestern Mediterranean Sea in October 2020. The mooring was near the site of neutrino telescope KM3NeT/ORCA (Adrián-Martinez et al., 2016) off the coast of Toulon, France, just 10 km south of the steep continental slope (and 5 km from its foot in the abyssal plain).

Fig. 1b shows the numbering of 45 vertical mooring lines, which were ordered in six groups for synchronisation purposes. As with all NIOZ4 T-sensors (van Haren, 2018), the individual clocks were synchronised via induction to a single standard clock on the mooring-array every 4 hours, so that all T-sensors were sampled within 0.01 s. Three buoys also held a Nortek AquaDopp single-point acoustic current meter. Details of design, construction and deployment are given in van Haren et al. (2021).

With the help from Irish Marine Institute Remotely Operated underwater Vehicle (ROV) "Holland I" on board Dutch R/V Pelagia, all lines with T-sensors were successfully cut and recovered in March 2024. Of the 45 lines, 43 were in good mechanical order, line 1.8 (line 8 of synchronization group 1; henceforth indicated without decimal point) was hit by the drag parachute whereby 10 sensors were lost, and line 65 was about 0.5 m lower than nominal because of a loop in the vertical line near the cable grid. Only line 36 did not register synchronisation, possibly due to an electric wire failure. Three T-sensors leaked and <10 were shifted in position due to tape malfunctioning. In total 2902 out of 2925 T-sensors functioned as expected mechanically.

Due to unknown causes all T-sensors switched off unintentionally when their file-size on the 8-GB Kingston memory card reached 30 MB. It implied that a maximum of 20 months of data was obtained, which were recorded at an interval of once per 2 s. After post-processing, 50-150, depending on moment in the record and type of analysis, extra T-sensors are not further considered due to electronics, noise problems. With respect to previous NIOZ4 version, here named 'NIOZ4o', the slightly modified electronics resulted in about twice lower noise levels of 0.00003°C and twice longer battery life.

As detailed elsewhere (van Haren, 2018), laboratory-bath calibration yielded a relative precision of <0.001°C. Instrumental electronic drift of typically 0.001°C mo$^{-1}$ after aging was primarily corrected by referencing daily-averaged vertical profiles, which must be stable from turbulent-overturning perspective in a stratified environment, to a smooth polynomial without instabilities. In addition, because vertical temperature (density) gradients are so small in the deep Mediterranean, so that buoyancy frequency $N = O(f)$ where f denotes the inertial frequency, reference was made to periods of typically one-hour duration that were quasi-homogeneous with temperature variations smaller than instrumental noise level (van Haren, 2022). Such >125-m tall quasi-homogeneous periods existed on days 350 (in 2020), 453, and 657 (-366 in 2021) in the records. This secondary drift correction allowed for proper calculations of turbulence values using the method of Thorpe (1977) under weakly stratified conditions. As will be demonstrated in Appendix A, under very weakly stratified conditions a tertiary correction involved low-pass filtering of data. This additional correction addresses short-term drift that was about 2-3 times larger in NIOZn than in NIOZo T-sensors.

**3 Results**

While a single-line mooring attached to an anchor at the seafloor may modify the nominal positions of instrumentation due to buoyancy-stretch of typically 0.1-1% of the total line length, depending on amount of buoyancy and line fabric and make, the large-ring mooring will experience an additional differential positioning due to the variable anchor height of the cable grid. Given the anchor being the steel ring, the distributed buoys on top of each of the 45 lines will pull the steel-cable grid (Fig. 1) up in the form of a dome. Hence, different heights are expected for different lines above the, presumed flat, seafloor.

Prior to deployment, stretch-tests were performed with the grid's steel cable. Under nominal tension of the planned buoyancy it was found that the exerted tension delivered a cable stretch so that the typical angle of grid inflection was expected to amount 5°. This angle to the horizontal could not be verified from visual inspection using ROV, although at various images a non-zero angle is discernible, which also seems to vary between different cable sections in the grid (Fig. 2a).

Eight 'corner-lines' (Fig. 2b) were also displaced to an amount not precisely verifiable from ROV-inspection. These lines were attached differently to the steel-cable grid as they could not be attached directly to their intersection points at the steel ring. Estimating the height of corner-line can be attempted from Fig. 2b. Visually, the right side of the small ring holding a vertical line in its center does not touch seafloor and the small ring rotates around the smallest of three short assist cables of which the centre is elevated to approximately $h = 0.6$ m above seafloor. If angles are measured on the basis of vertical/horizontal ratio 4/4.5, then the small-ring makes an angle of about 40° to the horizontal, so that its center is 0.8 m above the edge of the small-ring. In that case, the corner-line will start at $h = 1.0$ m.

**3.1 Parabola model**

Taking the 5°-angle due to distributed tension stretch as starting point, some simple models of (half-) cable-grid cross-section can be made (Fig. 3). Quasi-parabola and straight-cable models are considered. Considering that grid attachments are made in the center of the large-ring pipes at 0.3 m, the models start from there.

The simplest, albeit unrealistic, straight-cable model makes a fixed angle of 5° (green model in Fig, 3). A 5-m discretized parabola model intersecting the straight-cable model halfway will have its top at h = 2.07 m above (blue model). The steepest part, exceeding 5°, of the cable in this model has a horizontal distance of 5 m to the large ring, which corresponds with the position of, e.g., line 57 (cf. Fig. 1b). In the model, line 57 has its lowest T-sensor at h = 1.2 m. If an overall maximum angle of 5° is maintained, the top of that parabola model will be at h = 1.12 m, and the first line inside the large ring will have its lowest T-sensor at h = 0.72 m (red model).

In an attempt to verify these cable-grid models, the altimeter, and in a relative sense the pressure gauge, of the ROV gave a value of h = 0.7±0.1 m after landing on the small-ring of the first line of the grid's centre cable, pushing it to the seafloor from the low side. If the maximum-5° parabola model is correct, all vertical lines are a maximum of 0.4 m, or ±0.2 m, apart vertically. This is difficult to correct for in practice.

Unfortunately no pressure sensors were available, the three mounted on current meters being too inaccurate, to quantify height variations between lines. As a result, quantification is sought using the T-sensor data to verify, and possibly improve when necessary, above model values.

## 3.2 Adiabatic lapse rate height-determination method

Considering that the T-sensors have a noise level of 0.00003°C, potential temperature differences of >0.00005°C are statistically significantly detectable, in theory. Thus, given local deep Western Mediterranean adiabatic lapse rate of $\Gamma$ = -0.00017 °C m$^{-1}$ (here for simplicity a pressure of 10$^4$ Pa is transferred to a vertical distance of 1 m), vertical height differences of >0.3 m are potentially detectable using T-sensor data under near-homogeneous conditions in which temperature variations are predominantly due to compressibility effects. Such conditions do occur in the deep Western Mediterranean regularly, see the lower 250 m above seafloor in a shipborne-CTD profile (Fig. 4). $\Gamma$ dominates the temperature lapse with the vertical in Fig. 4a. In time series from moored T-sensor data, near-homogeneous conditions over 125-m vertical range occur about 60% of the time (Fig. 5a). These conditions lead to very low temperature variance across all frequencies outside instrumental white noise (Appendix A).

However, a complicating factor in 'adiabatic lapse rate height-determination' method is the electronic drift of T-sensors, which varies in intensity per sensor but typically amounts about 0.001°C mo$^{-1}$. While the value is one order of magnitude larger than the adiabatic lapse rate per unit length, in principle T-sensors attached to a particular vertical line can be corrected to within a precision (relative accuracy) of 0.0001°C (van Haren, 2018). All depends on a calibration with a precision of <0.0005°C, which is achievable using a thermostatic bath with constant temperature levels to within ±0.0001°C of their preset values. The standard post-processing correction is by fitting a smooth curve over sufficiently time-averaged vertical temperature profile that must be stable over an inertial period. When the temperature range is not too large, above precision is obtainable with some effort and careful search in the data. In the weakly stratified deep Mediterranean however, this correction is not achievable between different lines, because the low precision is not transferrable to a low absolute accuracy.

As a result, the height determinations from translated temperature differences between lines attached to the steel-cable grid are too large and erratically distributed (Fig. 6). Obviously, there is no consistency between lines in the image of Fig. 6, which shows no signs of expected lower (relative) values at the edges close to the ring and higher values near the center, following parabola models as in Fig. 3. Moreover, the variation in values is nearly an order of magnitude greater than expected from the parabola models in Fig. 3. Clustering per calibration -- approximately 190 T-sensors are used in the thermostatic bath per cycle (van Haren, 2018) -- does not give improvement of consistency in the image (not shown).

**3.3 Turbulence variance height-determination method**

As the adiabatic pressure effect on temperature is not a suitable measuring method for the expected doming of the steel-cable grid inside the large ring, another method is sought. Unexpectedly, such a method is not found during a near-homogeneous period. Instead, it is found when vertical temperature stratification is rather large, with peaks resulting in $N = 6f$, and turbulent temperature variations are large (Figs 5a, 7, 8). The combination of these two conditions, relatively large stratification and large turbulence, seems counter-intuitive, as stratification is generally considered to suppress turbulence.

Whilst stratification indeed suppresses the vertical length-scale of turbulence, it may have variable effects on temperature variance.

In case of the deep Western Mediterranean, relatively large vertical temperature gradients of a few 0.001°C over O(10-100) m occur with the advection of warmer waters (Figs 5a, 7a). The advection is possibly slanted towards the vertical, either induced by internal-wave action and/or by (sub-)mesoscale eddies. When mainly governed by planetary vorticity deflection, it represents in part convection-turbulence that appears in a vertically stratified environment at mid-latitudes (Marshall and Schott, 1999). All warming events observed thus far associate with considerable turbulence. In the entire time series (Fig. 5a) no significant cooling events occur. Current speeds (Fig. 5b) seldom exceed 0.1 m s$^{-1}$ and thus do not evidence strong flow events such as associated with deep dense-water formation that might occur in late winter, but is not observed.

Due to the relatively low-noise T-sensors, deep Western-Mediterranean waters can be characterized by frequency ($\omega$) spectra in which turbulence manifests itself over a range of at least two orders of magnitude (Fig. 8), approximately across $10 < \omega < 3000$ cpd (cycles per day), under the relatively large turbulence conditions. Outside this band, spectra are dominated by internal waves, for $\omega < 10$ cpd, and roll-off to instrumental white noise, for $\omega > 3000$ cpd. A strong temperature gradient produces high-frequency internal waves, but is also accompanied by turbulent eddies, probably as a result of breaking internal waves.

Here, we take the high-frequency portion $\Theta'(t, z)$ of the well-resolved turbulence band and, somewhat arbitrarily, band-pass filter between $600 < \omega < 1800$ cpd that is certainly outside internal wave and white noise bands. Although temperature variations in this range are part of inertial subrange reflecting a continual transfer between large energy-containing turbulence scales and small dissipative scales via shear-induced motions, such are mainly found well away from the seafloor. Within O(10) m from the seafloor, motions are predominantly of convection-turbulence nature in a buoyancy subrange, which manifests at all heights in the range $10 < \omega < 100$ cpd, albeit the spectral smoothing is coarse. Future investigations will be directed to improve statistics in part by averaging data from the 45 lines and data from different periods of stratified turbulence.

During such a period of warm waters from above, temperature variance may be relatively low closest to the seafloor, but it increases to high levels well above common interior-values in the first O(10) m above seafloor (Figs 8, 9). In this example, the peak in turbulence-temperature variance is found around h = 11 m. Above and below the peak one can take advantage of two depth-levels of high gradients in turbulence temperature-variance. Common interior-values are reached at about h > 40 m = $h_{sst}$, which could reflect the upper limit of layer of strong stratified turbulence 'sst' (Figs 7-9).

After scaling local turbulence temperature variance $\Theta'^2(z)$ with the 45-line average <.> value of its vertical gradient $d<\Theta'^2>/dz$ over dz = 2 m, a transfer from temperature to height value is established. Subsequently, below or above the peak value in Fig. 9, a height pattern can be computed relative to values of an arbitrary vertical line, 44 in this case (Fig. 10). Here, the pattern is given for height determination by computing across the largest gradient of temperature-variance, between T-sensors #2 and #3 from the seafloor.

The difference between this pattern and that in Fig. 6 is obvious. First, all values are between 0 and 2 m in Fig. 10, and a consistent statistical significance is found to within ±0.2 m. Cross-sections of the cable grid also confirm the doming of the pattern (Fig. 11). While the observed doming is close to the parabola models of Fig. 3, larger height-determination values than in the models are observed in the center, with slightly steeper overall grid cables that still roughly obey the maximum 5° slopes (Fig. 11). The ±0.2-m error range is easily verifiable after comparison with the provided bar. Corrections to vertical positioning of T-sensors are therefore feasible and necessary, because the difference between the center and edges of the cable grid is approximately 1.5 m.

**3.4 Stratified turbulence quantification**

The temperature variations of the well-stratified day 485 demonstrating the necessary height determination for the doming of the steel-cable grid show a background value of turbulence dissipation rate $O(10^{-10})$ m² s⁻³. Reduced values $O(10^{-11})$ m² s⁻³ are basically only found within a few meters from the seafloor, underneath the largest 2-m small-scale stratification with maximum buoyancy-frequency values of $N_{max} \approx 1.6\times10^{-3}$ s⁻¹ (Fig. 7). This is observable in time-depth plots of temperature, small-scale stratification and non-averaged turbulence dissipation rate 'values'. Coarsely every two hours, 124-m vertically averaged turbulence dissipation rate peaks, indicating the largest overturns being about 100 m in height, given a waterflow speed of 0.03 m s$^{-1}$. Turbulent overturns reach close to the seafloor, but only sporadically touch it, mostly at begin and end of the warm-water depression. The h = 40 m of elevated high-frequency temperature variance (Fig. 9) and stratification (Fig. 7b) show non-negligible turbulence dissipation rate values with further elevated values reaching the seafloor before and after the warm-water passage (Fig. 7c).

Time-depth mean values from the 1.3-day period are for turbulence dissipation rate $[<\varepsilon>] = 6\pm3\times10^{-10}$ m$^2$s$^{-3}$ and for turbulent diffusivity $[<K_z>] = 1.4\pm0.7\times10^{-3}$ m$^2$s$^{-1}$ under $[<N>] = 2.8\pm0.3\times10^{-4}$ s$^{-1} \approx 3f$. These 1.3-day, 124-m mean turbulence values are about one order of magnitude larger than open-ocean values observed in stratified waters well away from boundaries (Gregg 1989; de Lavergne et al. 2020; Yasuda et al. 2021).

Although the warm-water event of Fig. 7 is relatively strong, it is not exceptional and elevated temperature-stratification and -variance alternate in time with near-homogeneous episodes throughout the 20-month records (Fig. 5a). This will be reported elsewhere in more detail, notably using three-dimensional investigations.

4 Conclusions

The expected height variation due to vertical buoyancy pull across a steel-cable grid, which was suspended within a large anchoring steel-pipe ring, was modelled as an inverted parabolic with maximum 5° angle to the horizontal. To verify the height variation of the instrumented vertical mooring lines across the grid, we expected to use an observational period with unmeasurably small vertical density (temperature) stratification so that the adiabatic lapse rate $\Gamma = dT/dz$ would dominate vertical temperature variations T(z). The T-sensors have a noise level of about 0.00003°C, while $|\Gamma| \approx 0.00017$°C m$^{-1}$ in the deep Western Mediterranean and thus is potentially measurable by sensors nominally 2-m apart vertically. However, the sensor's electronic drift at all scales turned out insufficiently correctable under near-homogeneous conditions.

Instead, a mooring-height determination was found during a period of relatively large stratification, during a slump down of warm water presumably slanted from above and induced by internal waves. By band-pass filtering the highest resolved turbulence variance, mainly from inertial subrange, across the strongest temperature gradient, the dome of pulled-up grid was significantly detectable, and fine-tuned a parabola model with height variations between the moorings of correctable (0.5-2.0)±0.2 m.

The impact of investigating turbulence signals from high-resolution moored T-sensors the deep Mediterranean is several-fold. First, it demonstrates the dynamics of internal-wave breaking governed by either near-inertial or sub-mesoscale motions slumping relatively warm waters to within a few meters above the seafloor. Thereby, an episodic-average turbulence dissipation rate is provided, which is about ten times larger than ambient values above a flat seafloor. The enhanced turbulence affects deep-sea life. Deep-sea turbulence is studied more elaborately in van Haren (2026 submitted).

Second, the strong vertical variation in turbulence temperature-variance profiles, across relatively large local vertical temperature gradients, may be useful for height determinations in nearby moorings over flat seafloors also in shallow seas, and, more difficult, above sloping seafloors, whereby a correction may be applied for (unknown) mooring-line stretch under tensioning by buoyancy. Such height determinations may also be necessary when moorings are accidentally placed on small rocks or in small gullies. The resulting determination demonstrated that the parabola model based on in-house line-tensioning was adequate and required only secondary adjustment in the slightly steeper cables of the underwater large-ring mooring, albeit all showed <5° sloping to the horizontal as anticipated from the in-house tests.

*Data availability.* Only raw data are stored from the T-sensor mooring-array. Analyses proceed via extensive post-processing, including manual checks, which are adapted to the specific analysis task. Because of the complex processing the raw data are not made publicly accessible. van Haren (2025): "Large-ring mooring current meter and CTD data", Mendeley Data, V1, https://doi.org/10.17632/f8kfwcvtdn.1.

*Competing interests.* The author has no competing interests.

*Acknowledgments.* This research was supported in part by NWO, the Netherlands organization for the advancement of science. Captains and crews of R/V Pelagia are thanked for the very pleasant cooperation. The team of ROV Holland I performed an excellent underwater mission to recover the instrumentation of the large ring. NIOZ colleagues notably from the NMF department are especially thanked for their indispensable contributions during the long preparatory and construction phases to make this unique sea-operation successful. I am indebted to colleagues in the KM3NeT Collaboration, who demonstrated unison to get large-scale infrastructural projects funded. M. de Jong and A. Heijboer helped in securing NWO funding.

**Appendix A Extra drift correction for T-sensors in near-homogeneous waters**

When waters are very weakly stratified or near-homogeneous over the range of moored T-sensors, a short-term drift error may emerge. This drift partially causes the impossibility to determine instrumental height variations under such conditions. Albeit electronic drift is well known to occur on long timescales of weeks-months, short-term hourly drift may appear because of nonlinear temperature dependency and/or inadequate contact between the NTC's and the environment through the glass tube and its contact paste. Negative Temperature Coefficient 'NTC' thermistors are the measuring component of NIOZ T-sensors. This short-term drift was previously observed in NIOZ4o deep-trench data and, especially clear, in air (van Haren and Bosveld, 2022). It turned out difficult to correct for. During a 2017/2018 test experiment in the deep Western Mediterranean it did not pose a great problem in NIOZ4o data. Unfortunately, NIOZ4n appear to have about twice larger short-term drift than previous NIOZ4o, despite their smaller long-term drift and smaller instrumental noise, both by a factor of two-three approximately. A correction for short-term drift of NIOZ4n is proposed below, with reference to NIOZ4o.

Half-day NIOZ4n data (Fig. A1a-c) from arbitrary line 25 are compared with a 104-m tall set of NIOZ4o data (Fig. A1d-f). The investigated samples are from almost homogeneous waters, with a total colour-range over a Conservative-Temperature difference of only 0.00017°C. Although NIOZ4o are more noisy (Fig. A1a,d), the NIOZ4n show a more horizontal-stripy pattern that has different values through time, compared to T-sensors above and below. This is evidence of remaining bias due to short-term drift. Low-pass time filtering does not reduce these (Fig. A1b,e), but additional vertical filtering adequately removes the bias (Fig. A1c,f).

As a consequence of near-homogeneity, energetic overturning scales are expected to be large due to the reduced restoring force. In both data sets in the center of images, albeit clearer in the NIOZ4o, one notices a slanting jet of warmer waters over a vertical range of about 60 m in short bursts of 10-20 m. Such jets of convection turbulence were found abundant in the Mariana Trench (van Haren, 2023), but relatively rarely in the deep Mediterranean.

Depending on the rate of stratification, the vertical filter cut-off at 0.05-0.2 cpm (cycles per meter) is obtained after fine-tuning in an attempt to retain the relevant overturning scales as much as possible.

Under weakly but stable stratified conditions 0.1-0.2 cpm is used, while under near-homogeneous and unstable conditions 0.05-0.1 cpm is used. The fine-tuning of the vertical filter concerns relatively adequate spectral improvement and turbulence calculations. Naturally, all data-corrections yield a certain loss of information, but it is informative to estimate how much the loss may be.

In 4-d-average spectra (Fig. A2) that include the 0.5-d period of Fig. A1, the impact of small-scale drift is seen to be larger for NIOZ4n (Fig. A2a) than for NIOZ4o (Fig. A2b). In these plots spectra are scaled with the slope of buoyancy subrange, for clarity. As a reference for the correction, the temperature difference $\Delta\Theta$ (magenta spectrum) is taken between the two neighbouring T-sensors at h = 29 and 31 m. That difference spectrum is compared with the spectrum of temperature data from the upper T-sensor (green). The magenta spectrum has a higher noise level by about a factor of two for $\omega > 100$ cpd than that of the green spectrum. This is commensurate with random white noise.

At lower frequencies, the magenta spectrum crosses the green spectrum around 50 and 150 cpd for Fig. A2a and A2b, respectively. This means that data are no longer dominated by white noise, but by other parametrizations, which are governed either by natural processes or by instrumental flaws other than noise. Around the crossing frequency, temperature spectra become horizontal following buoyancy-subrange scaling, rather abruptly. This scaling, which represents convection turbulence of an active scalar (Bolgiano, 1959; Obukhov, 1959), was no longer dominant at frequencies higher than that of the crossing, more so in Fig. A2a than in Fig. A2b.

After correction by applying vertical low-pass-filtering (red), the weak slope towards lower frequencies (especially that of Fig. A2a) is correctly removed and white noise levels are lower. The spectral slope change to white noise is now at the same frequency, 400 cpd, for both data sets. The quasi-transfer function of correction depicted in the blue spectra is less steeply sloping for $\omega < 1000$ cpd in Fig. A2a than in Fig. A2b, which demonstrates the larger effects of short-term drift correction for the NIOZ4n compared to the NIOZ4o. However, in both data sets of very weakly stratified deep-sea waters it prevents resolution of the transition from buoyancy and/or inertial subranges to the viscous turbulence dissipation range.

Resuming, after vertical-filtering correction, temperature data at ω < 400 cpd seem useful for turbulence calculations under weakly stratified conditions. Note that this correction is not needed during periods with relatively large temperature variance and stratified, generally more shear-induced, turbulence. For turbulence dissipation rate calculations, 10-30% reduction is obtained from short-term drift correction. This reduction is well within the error range of a factor of two normally achieved for ocean turbulence data.

# References

Adrián-Martinez, S. et al.: Letter of intent for KM3NeT 2.0, J. Phys. G, 43, 084001, 2016.

Bolgiano, R.: Turbulent spectra in a stably stratifed atmosphere, J. Geophys. Res. 64, 2226-2229, 1959.

de Lavergne, C., Vic, C., Madec, G., Roquet, F., Waterhouse, A. F., and Whalen, C. B. et al.: A parameterization of local and remote tidal mixing, J. Adv. Mod. Earth Sys., 12, e2020MS002065, 2020.

Gregg, M. C.: Scaling turbulent dissipation in the thermocline, J. Geophys. Res., 94, 9686-9698, 1989.

IOC, SCOR, and IAPSO: The International Thermodynamic Equation of Seawater – 2010: Calculation and Use of Thermodynamic Properties, Intergovernmental Oceanographic Commission, Manuals and Guides No. 56, UNESCO, Paris (F), 196 pp, 2010.

Marshall, J., and Schott, F.: Open-ocean convection: Observations, theory, and models, Rev. Geophys., 37, 1-64, 1999.

Obukhov, A. M.: Effect of buoyancy forces on the structure of temperature field in a turbulent flow, Dokl. Akad. Nauk SSSR, 125, 1246-1248, 1959.

Thorpe, S. A.: Turbulence and mixing in a Scottish loch, Phil. Trans. Roy. Soc. Lond. A, 286, 125-181, 1977.

van Haren, H.: Philosophy and application of high-resolution temperature sensors for stratified waters, Sensors, 18, 3184, doi:10.3390/s18103184, 2018.

van Haren, H.: Thermistor string corrections in data from very weakly stratified deep-ocean waters, Deep-Sea Res. I, 189, 103870, 2022.

van Haren, H.: How and what turbulent are deep Mariana Trench waters? Dyn. Atmos. Oc., 103, 101372, 2023.

van Haren, H., and Bosveld, F. C.: Internal wave and turbulence observations with very high-resolution temperature sensors along the Cabauw mast, J. Atmos. Ocean. Technol., 39, 1149-1165, 2022.

van Haren, H., Bakker, R., Witte, Y., Laan, M., and van Heerwaarden, J.: Half a cubic hectometer mooring-array 3D-T of 3000 temperature sensors in the deep sea, J. Atmos. Ocean. Technol., 38, 1585-1597, 2021.

van Haren, H., et al.: Whipped and mixed warm clouds in the deep sea, Geophys. Res. Lett., in press, 2026.

Yasuda, I., et al.: Estimate of turbulent energy dissipation rate using free‑fall and CTD‑attached fast‑response thermistors in weak ocean turbulence, J. Oceanogr., 77, 17-28, 2021.

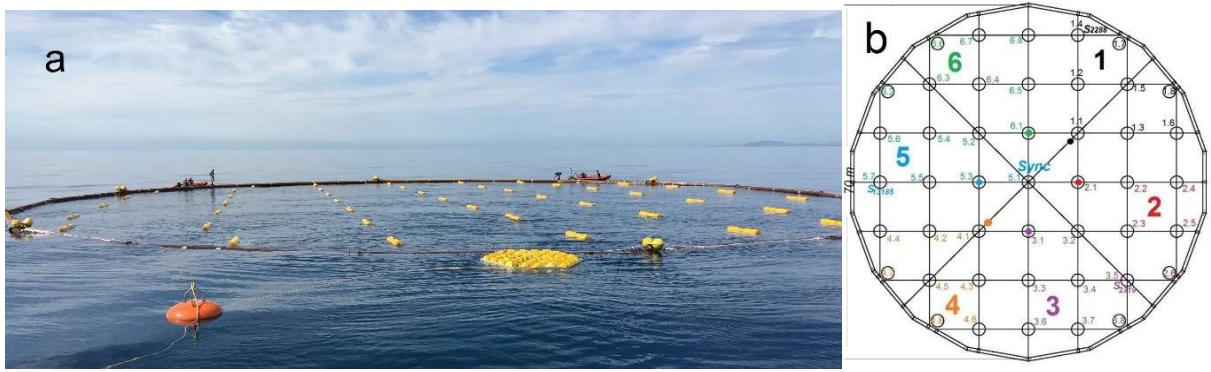

**Figure 1.** Large ring in fold-up form at sea, during deployment just prior to finish the opening of air-valves. The front part of the large steel-pipe ring is already underwater. Almost all top-buoys of the 45 small-ring-compacted mooring lines are visible. In the front still outside the ring, the yellow drag parachute and orange pick-up buoy are floating. (b) Layout of the large-ring mooring, with steel-cable grid and small-rings numbered in six synchronisation groups. Lines 14, 35 and 57 (omitting the decimal point) held a current meter at the top-buoy.

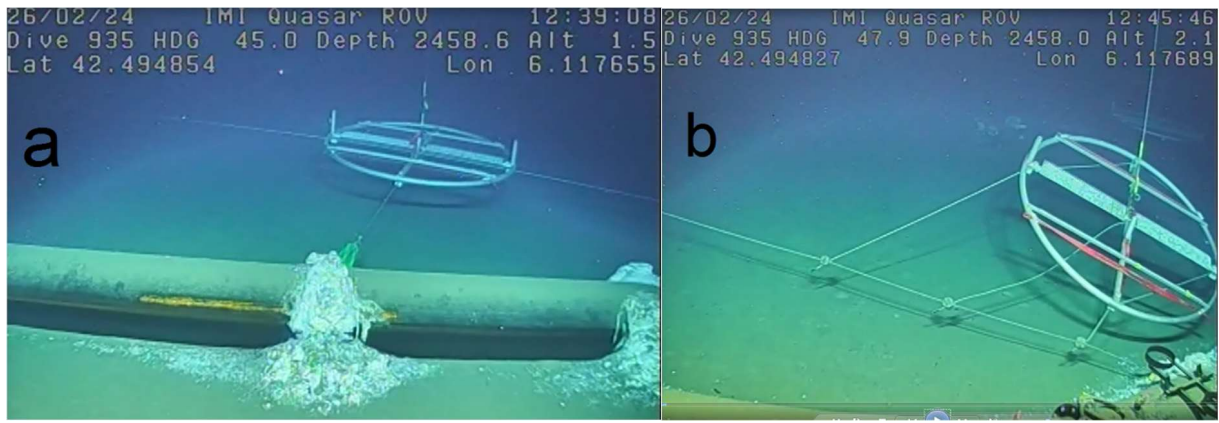

**Figure 2.** Underwater video stills of small-rings demonstrating steel-cable grid elevation and cable-inclinations. The 0.61-m diameter steel pipe in the foreground is part of the large anchoring ring, which sank 0.07±0.02 m in the sediment of the <1° flat seafloor. All steel cables are attached to the middle of the steel pipes, and thus at height h = 0.24 m above sediment. (a) Line 44 (cf. Fig. 1b). To the right of the small-ring the wire visibly makes a larger angle to the vertical than to the left. (b) Estimating height of 'corner-line' 47, see text. (Images from video by ROV Holland I).

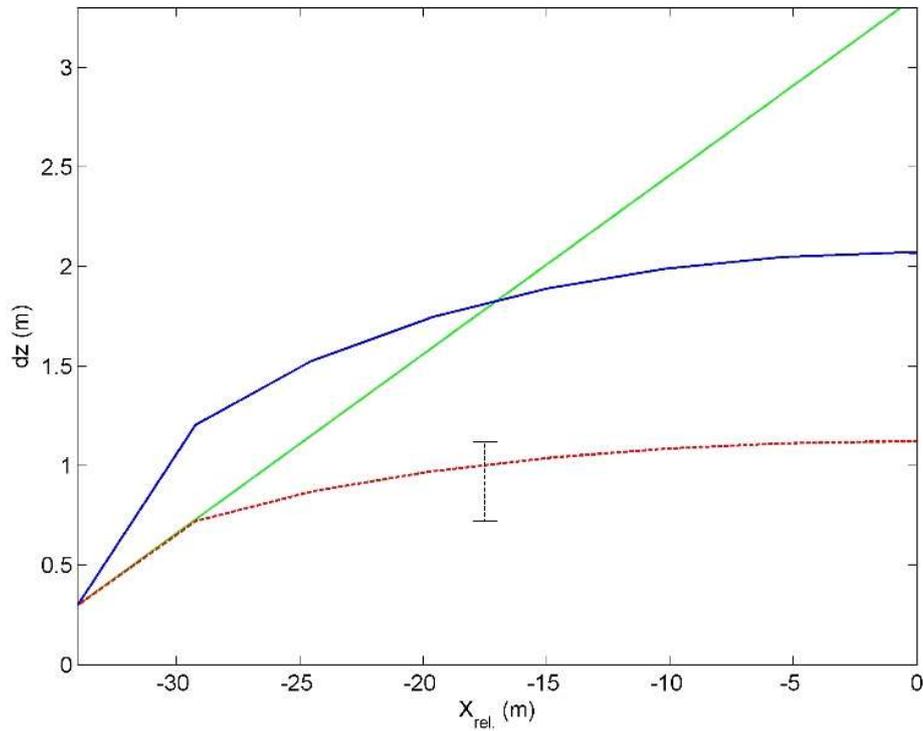

**Figure 3.** Quasi-parabola and straight-line models, for half centre steel cable of large-ring mooring grid. The seafloor is at the horizontal axis, the cable-grid attachment to the large-ring is for a steel pipe at a solid floor. The green straight line makes a fixed angle of 5° with the horizontal, which angle was established after in-port tension tests. The blue (solid line, 5-m discretized) parabola model intersects the green line halfway, so that its top is at h = 2.07 m. If an overall maximum angle of 5° is maintained (red dotted model), the top is at h = 1.12 m, and the first vertical line attached to relative horizontal position $x_{rel}$ = -29 m will be at h = 0.72 m.

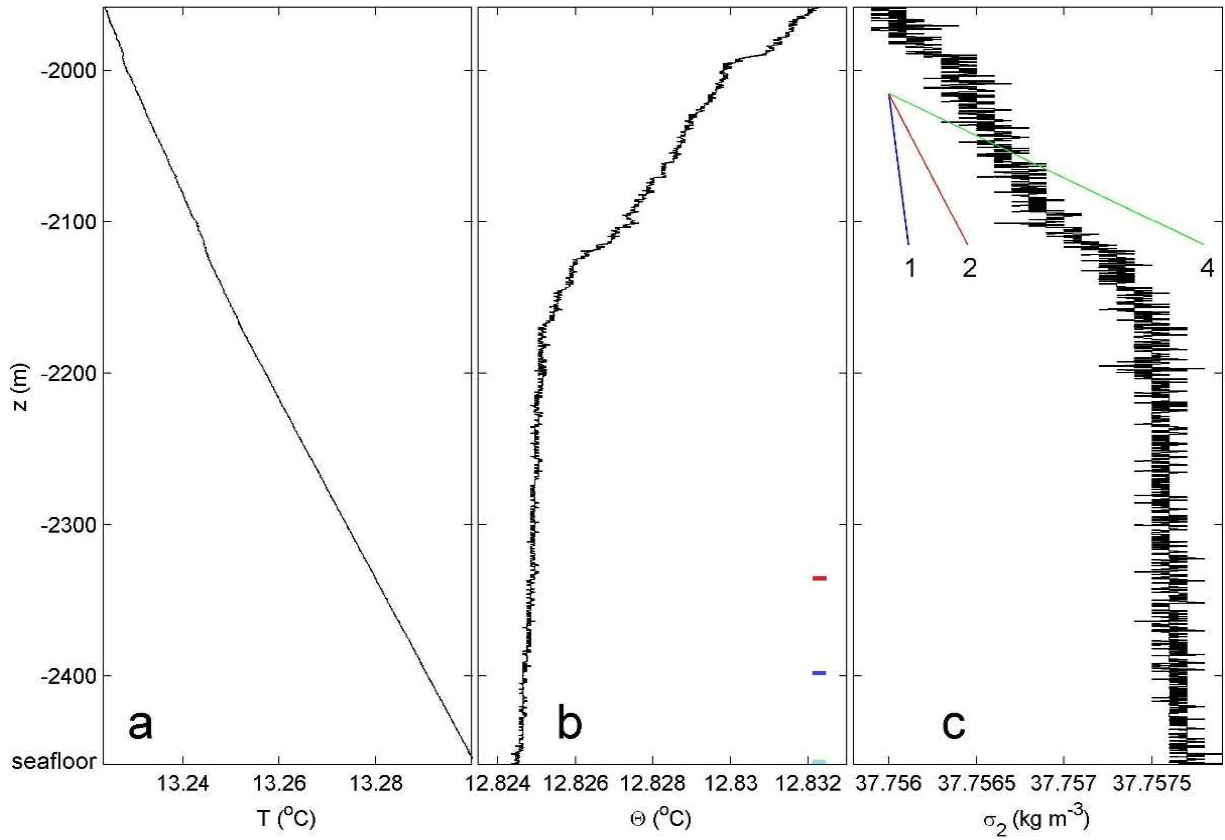

**Figure 4.** Lower 500 m of shipborne CTD-profile data obtained near the large ring during mooring deployment. (a) Uncorrected temperature. (b) Conservative Temperature (IOC et al., 2010) corrected for compression. Small colour bars indicate nominal heights of moored T-sensors at lowest (cyan), middle (blue) and upper (red) positions. (c) Density anomaly referenced to $2\times10^7$ Pa. The sloping lines indicate several stratification rates in terms of buoyancy frequency $N = xf$, $x$ = 1, 2, 4 times the local inertial frequency $f$.

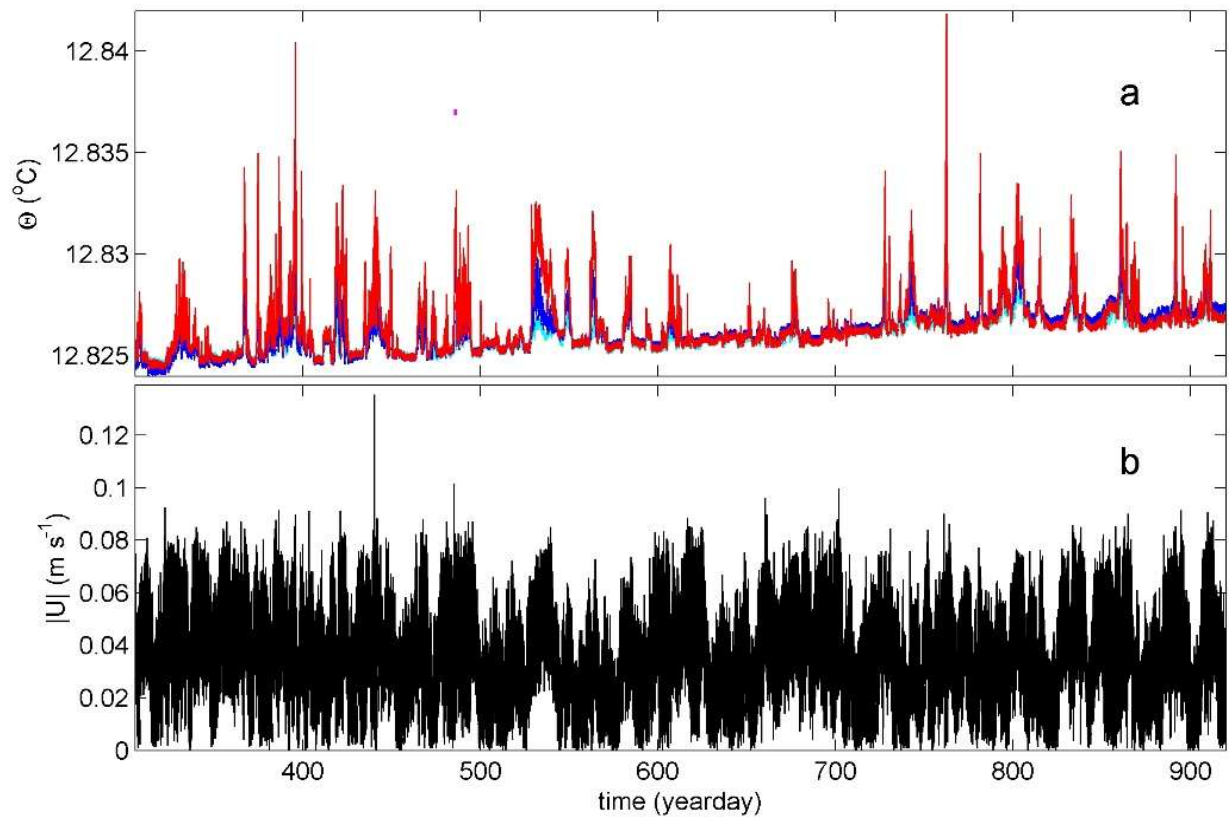

**Figure 5.** Overall 20-month time series of moored temperature and current meter data. (a) Conservative Temperature at h = 1 (cyan), 63 (blue) and 125 m (red) cf. Fig. 4b of arbitrary line 53. Data are not corrected for electronic-drift bias. The magenta dot indicates day 485. (b) Unfiltered current speed at h = 126 m of line 14.

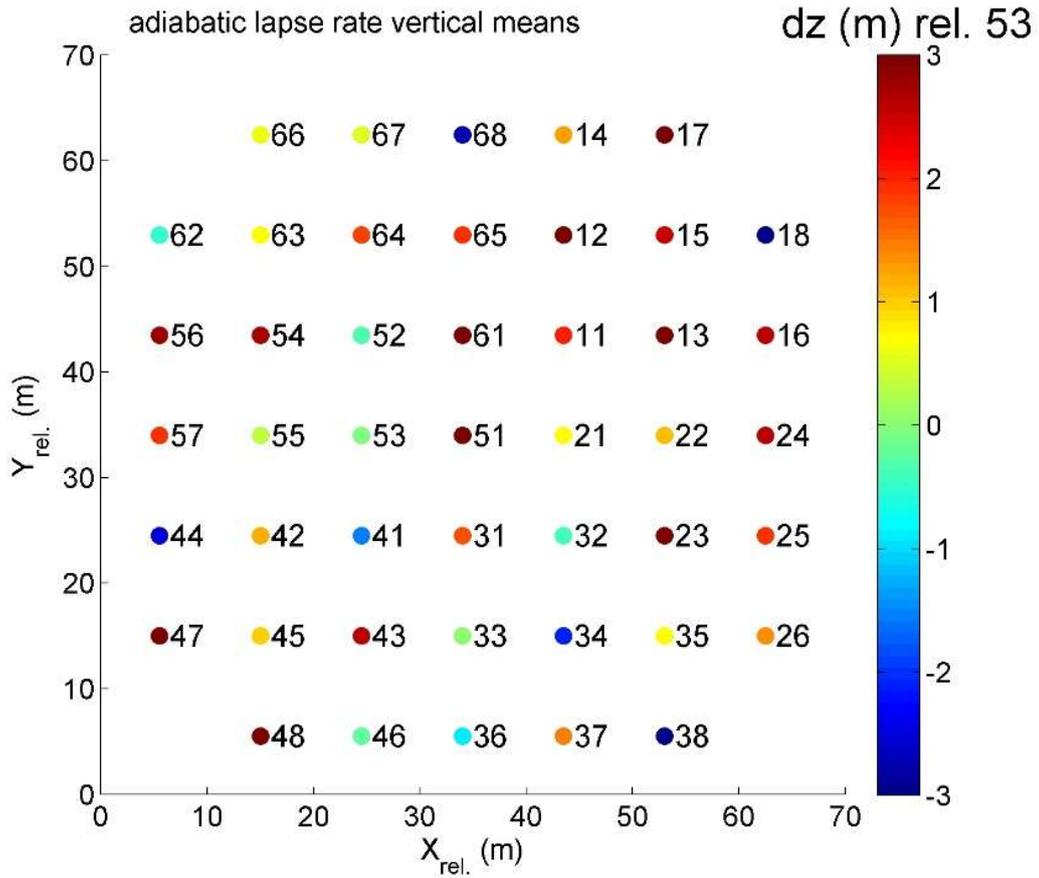

**Figure 6.** Vertical displacement calculation for vertical lines of the large-ring mooring using the local adiabatic lapse rate $\Gamma$. This height-determination method is based on temperature shift per line after calibration and drift correction for near-homogeneous period 350.04-350.08, with line 53 as reference. The conversion of meters into degrees Celsius is via $\Gamma = 0.00017°C\ m^{-1}$, so that a vertical difference of dz = 3 m reflects approximately 0.0005°C. Corner-lines are 17, 18, 26, 38, 47, 48, 62 and 66.

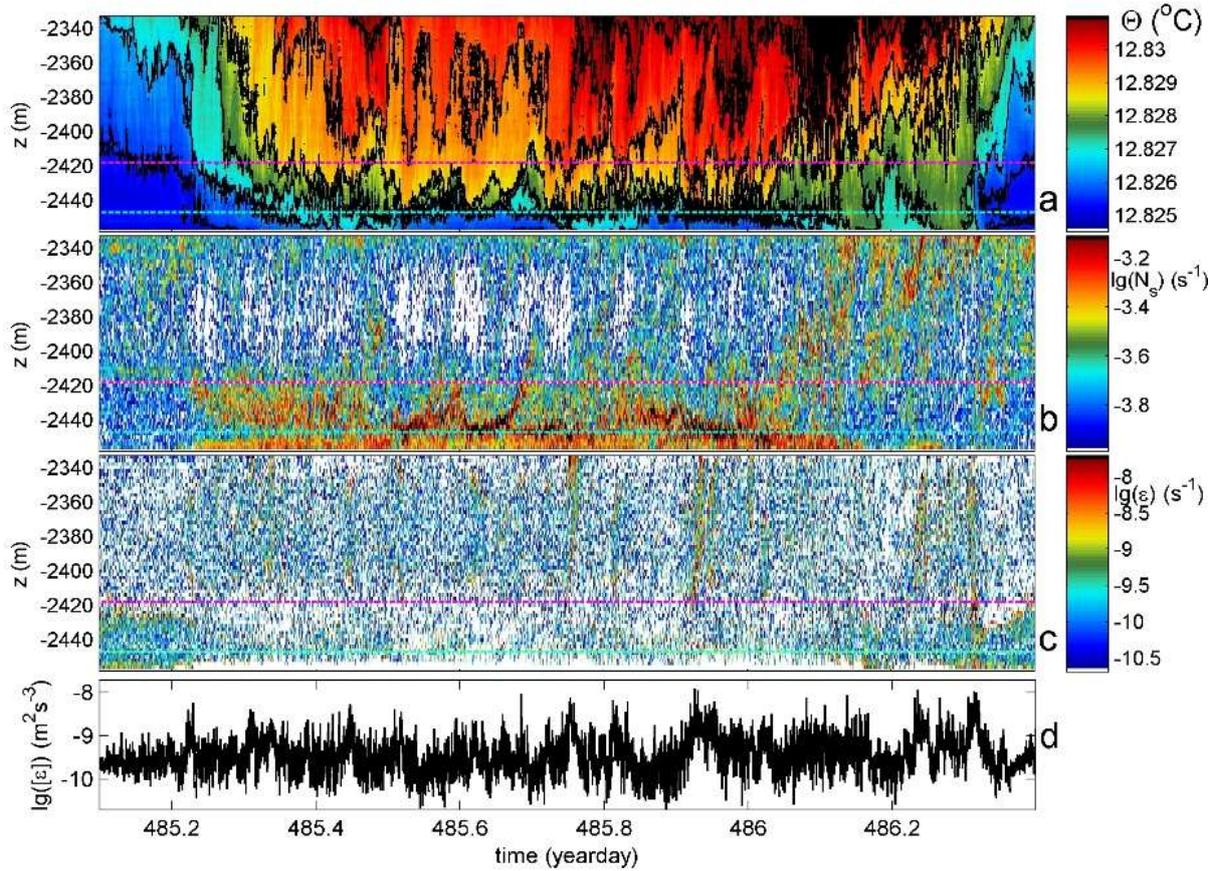

**Figure 7**. Thirty-one hours of data from line 53 during a turbulent passage of relatively warm water. Horizontal dashed magenta and cyan reference lines are at h = 40 and 11 m above seafloor, respectively. (a) Time-depth plot of Conservative Temperature with black contours every 0.001°C. (b) Logarithm of 2-m small-scale buoyancy frequency from reordered profiles of data in a. (c) Logarithm of non-averaged turbulence dissipation rate from data in a. (d) Time series of logarithm of turbulence dissipation rate averaged over the 124-m vertical extent of T-sensors.

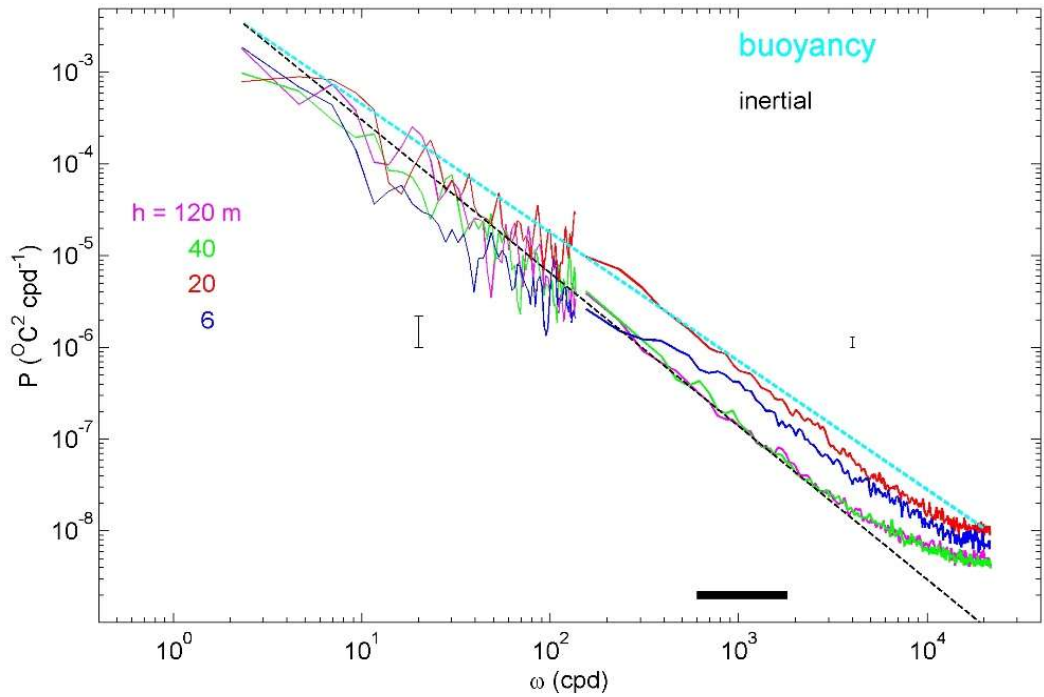

**Figure 8.** Unscaled frequency spectra, patched from weakly and heavily smoothed parts, for four T-sensors of line 53 at indicated heights h above seafloor, averaged over the 1.3-day period of Fig. 7. Spectral slopes for inertial and buoyancy subranges (of turbulence) are indicated with straight dashed lines. The horizontal black bar indicates the pass-filter band that is applied for the turbulence-variance method of height determination.

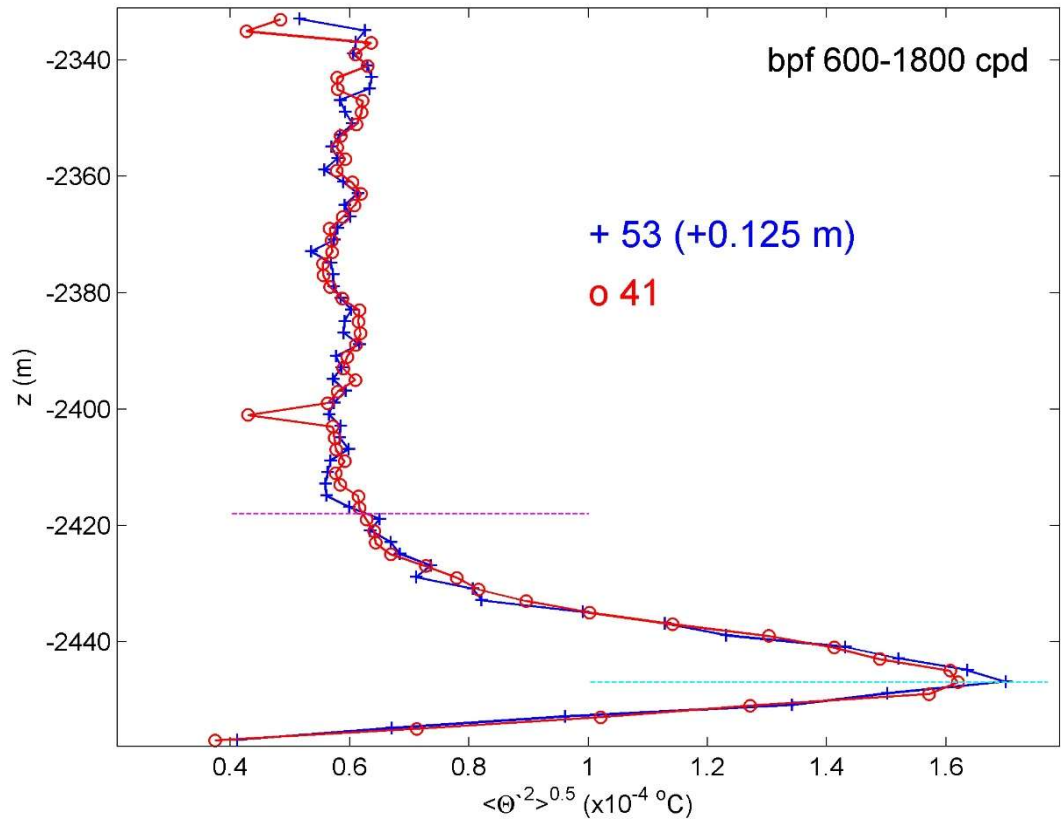

**Figure 9.** Vertical profiles of standard deviation of band-pass filtered 'bpf' high-frequency turbulence signals for temperature data in Figs 7a, 8 for two neighbouring lines, with off-set relative height determination. The variance-peak height (cyan-dashed line) corresponds also to the height of strongest layering in stratification (sic!) in Fig. 7b. The magenta-dashed line delineates the vertical extent of enhanced temperature variance above interior values.

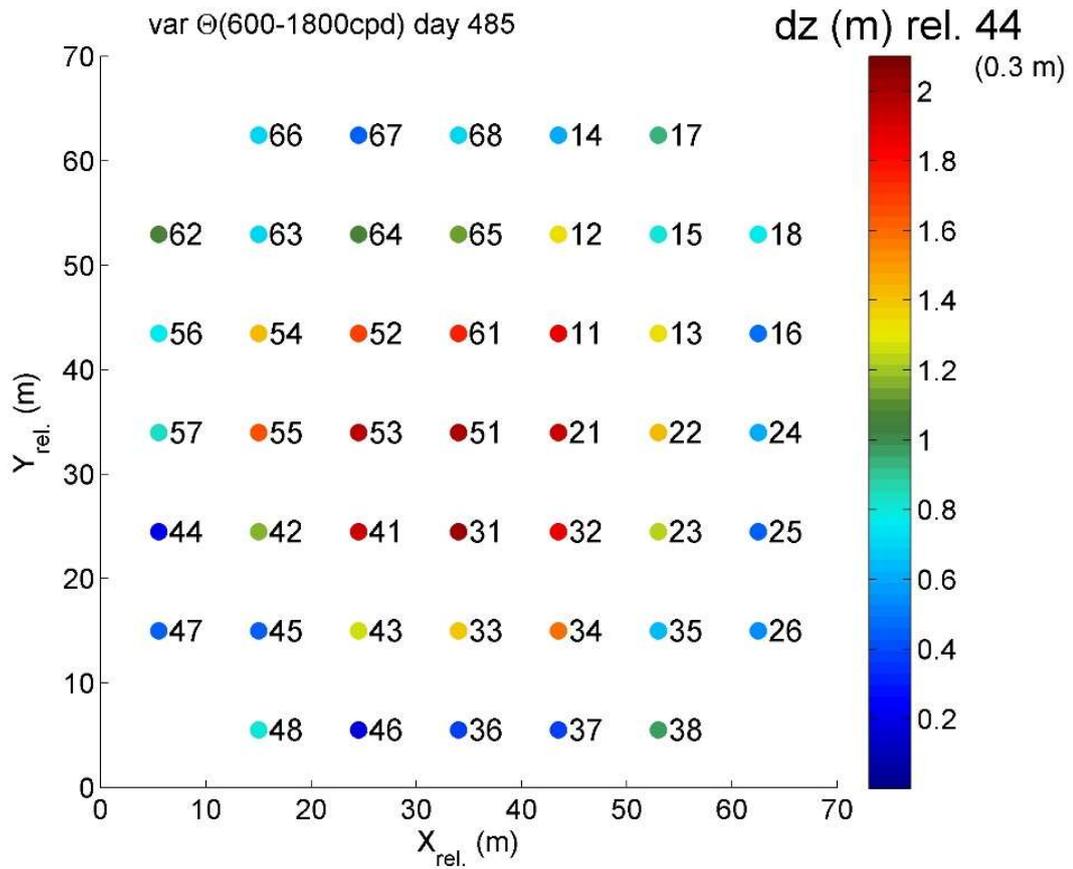

**Figure 10.** As Fig. 6, but for turbulence-variance height-determination method determined from profiles like in Fig. 9 using T-sensor data between positions 2 and 3 above seafloor, where the gradient in temperature variance is maximum, divided by the average gradient over 2 m. Values are given relative to those of line 44 (0.3 m).

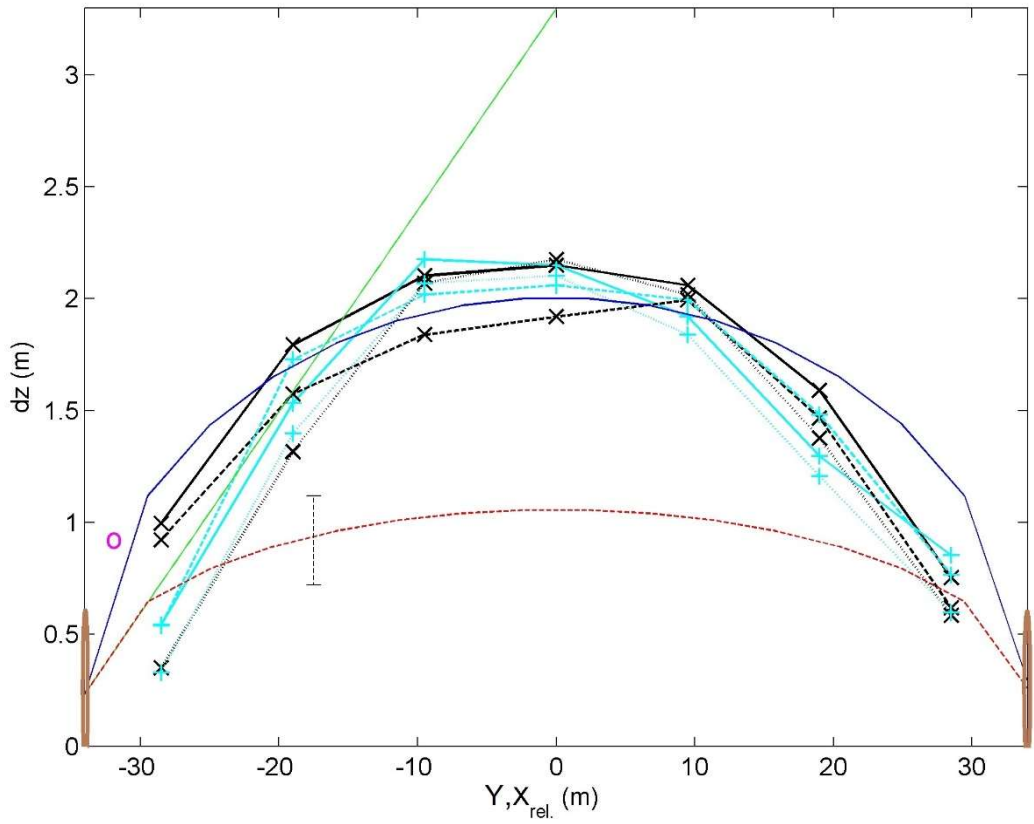

**Figure 11**. Constant-Y (black x graphs) and constant-X (cyan +) cross-sections without corner-lines of height determinations from Fig. 10. Solid lines indicate center lines in both directions. Corner-line height determination is indicated by magenta o. In the background, models are given in green, blue and red of double distance than in Fig. 3.

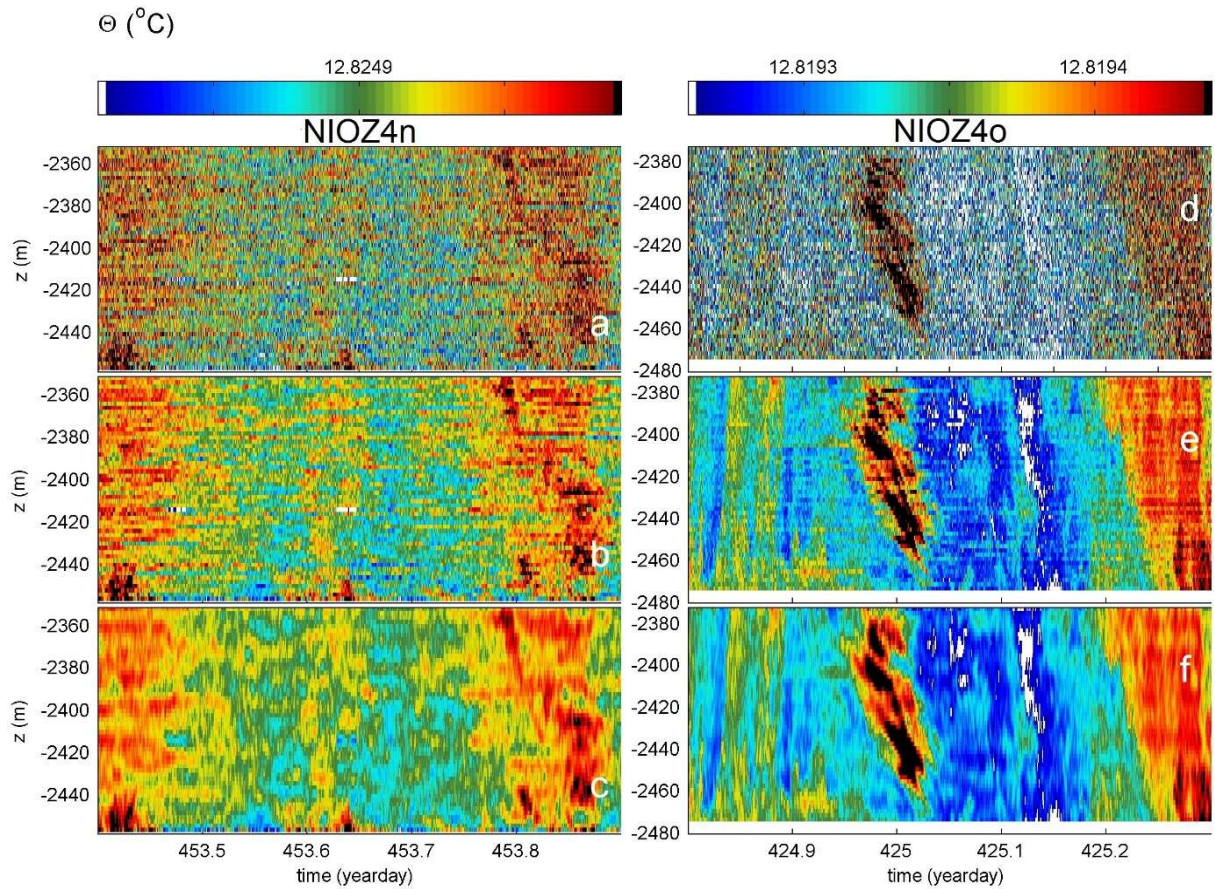

**Figure A1**. Half-day Conservative-Temperature data from h = 1-104 m demonstrating the correction of short-term drift. The conditions are near homogeneous, as full the temperature range is only 0.00017°C during the present experiment in 2020/2021 (left column) and during a test-experiment in 22-m deeper water in 2017/2018 (right column). (a, d). Unfiltered data, after post-processing involving calibration, referencing to CTD-, homogeneous-period-, and smooth-polynomial data. (b, e) Low-pass filtered 'lpf' with cut-off at 500 cpd. (c, f) Corrected for short-term drift: 500-cpd and 10-m vertical-scale lpf data.

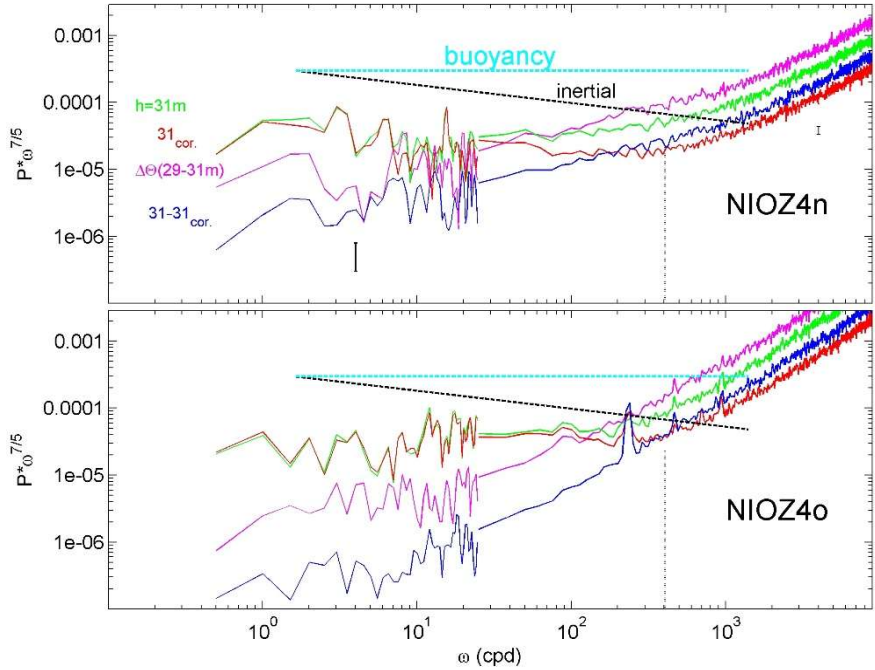

**Figure A2**. Four-day average spectra that are patched together from two, a weakly- and a heavily smoothed part, and scaled with the buoyancy-subrange slope (horizontal cyan). The spectra demonstrate effects of and correction for short-term drift in T-sensor data around h = 30 m during near-homogeneous periods between days 453-457, including those of Fig. A1. Plotted are spectra for unfiltered data (green), vertical temperature difference with data from T-sensor 2-m lower (magenta), 500-cpd and 10-m vertical scale corrected data (red), and the difference between green and red spectra (blue). For reference, the relative log-log plot slope is given for inertial subrange (black). The vertical dotted line at 400 cpd is explained in the text. a) NIOZ4n data. b) NIOZ4o data.